# Design of a millifluidic device for the evaluation of blue-light disinfection on *Pseudomonas aeruginosa* PA01 biofilm

Nidia Maldonado-Carmona,[a,c] Murielle Baltazar,[b,c] Joanne L. Fothergill,[b,c] Nelly Henry,[a,c] Light4Lungs group[c]

[a] Sorbonne Université, Centre National de la Recherche Scientifique, Laboratoire Jean Perrin, Paris, France
[b] University of Liverpool, Ronald Ross Building, Department of Clinical Infection Microbiology and Immunology, Liverpool, UK
[c] Light4Lungs group: S. Nonell, Institut Quimic de Sarria, Barcelona, Spain; G. Romano, University of Florence, Florence, Italy; N. Henry, Sorbonne Université, Paris, France; M. Palomares, Fundacio Privada Institut Catala D'Investigacio Quimica, Tarragona, Spain; F. Moret, University of Padova, Padova, Italy; G. Taylor, Cardiff Scintigraphics Ltd, Cardiff, UK; A. Kadioglu, The University of Liverpool, Liverpool, UK

## Abstract

**Significance:** For the first time to our knowledge, a millifluidic device has been designed to observe the effects of blue-light irradiation in *Pseudomonas aeruginosa* PA01 bacterial biofilm.

**Approach:** A bacterial biofilm is formed under controlled flow of nutrients in microfabricated channels settled on a microscope stage. The setup is devised to apply defined irradiation and evaluate the photo-killing efficiency *in situ* using video-microscopy and fluorescent probes.

**Results:** We investigated bacterial survival after delivering a precise, controlled and spatially defined blue light dose upon growing biofilm. We demonstrate in this report that the combined use of constitutive GFP and propidium iodide in a millifluidic setup enables to evidence light-induced mortality at the single cell and community level.

**Conclusions:** Real-time monitoring of a bacterial biofilm growing in a millifluidic device after irradiation opens an avenue for a better understanding of the biofilm community response to blue-light, permitting researchers to approach photokilling efficiency of biofilm as a dynamic process.

## 1 Introduction and Background

Pathogenic biofilms proliferation represents a burden for human health and a challenge for the development of antibiotic alternatives[1]. Blue-light irradiation intends to use bacterial tetrapyrrolic metabolites (i.e., protoporphyrin IX) as photosensitizers, together with specific blue-light irradiation, for reactiven oxygen species generation[2]. Millifluidic devices have already been used to describe the growth and complexity of biofilm communities[3], and, together with live/dead fluorescent labelling, it could be used to analyze the cellular and community response to blue-light irradiation. As part of the Light4Lungs initiative, the present work establishes for the first time the design of a millifluidic setup for biofilm irradiation.

## 2   Aims

Develop a millifluidic device for the growth and real-time observation of *Pseudomonas aeruginosa* PA01 biofilm development and response to *in situ* blue-light irradiation, using video microscopy and live/dead specific fluorescent probes.

## 3   Methods

*Bacteria strains, culture media and reagents:* Green-fluorescent *Pseudomonas aeruginosa* PA01-GFP[4] was provided by the University of Liverpool. Upon utilization a smear of the frozen stock was inoculated into M9 media[5] supplemented with casaminoacids 2 g/L and propidium iodide (PI) 3 µM (M9PI). *Device design and construction:* The device was constructed as elsewhere described[3]. Polydimethylsiloxane mixture (RTV615A+B 9:1 mixture, Momentive Performance Materials) was used to build a template with channels, which were then drilled and bound to glass coverslips using an oxygen plasma cleaner (Harrick) (Figure 1A). *Imaging and light irradiation setup (Figure 1B):* Samples were observed using an inverted NIKON TE300, equipped with motorized x, y and z displacements and shutters, equipped with a Hamamatsu ORCA-R2 EMCCD camera. Acquisitions: brightfield images were collected in direct illumination; PI channel, excitation: 550 nm, m-Cherry filter (Ex. 562/40nm, DM. 593, Em. 641/75); GFP channel, excitation: 490 nm, green filter for GFP filter (Ex. 482/35, DM 506 nm, FF01-536/40). Irradiation: lightsource: 405 nm laser diode (LDI-7, 89North); power: 0.0161 W (measured *in situ* with a powermeter (Thorlabs)); irradiation area: 0.00333 $cm^2$; irradiation time: 111 seconds; light dose: 536.7 $J/cm^2$. *Biofilm formation:* Overnight bacteria cultures (500 rpm, 37 ºC) were diluted to obtain an optical density of 0.1, and further incubated during two hours. The culture was diluted in 5 ml of M9PI (OD 0.1, ~$10^7$ CFU/mL). Bacteria were injected into the millifluidic device and incubated at 37 ºC for 30 minutes. After incubation, the channel was connected to a syringe pumping M9PI at a 0.5 ml/hr flow rate and acquisitions started (Figure 1C).

## 4   Results

Previous efforts made in our group have unveiled the utility of using millifluidic devices for monitoring biofilm communities in a controlled environment[3]. This device provides a constant supply of nutrients, and a water-biofilm interface, which needs to be considered when evaluating the biofilm's photokilling efficiency[6]. PA01-GFP biofilm development was observed close to the PDMS surface, and followed up during the different development stages, cells were seen as individual cells, growing to clusters and finally forming a continuous biofilm layer. Biofilm was irradiated at two hours (Figure 1D) and four hours of development in the channel (Figure 1E). Upon irradiation GFP fluorescence was dimed, due to photobleaching. Meanwhile, PI-labelled bacteria appeared in the red channel, indicating that PI is able to permeate through sensitized membranes and bind to nucleic acids inside the cell, a process closely correlated with bacterial death[7]. PI labelling reaches a maximum value after about one hour, suggesting membrane permeabilization due to blue-light induced damage is not an instantaneous process. (Figure 1D). At a community scale, the *in situ* irradiation is able to sensitize a precisely defined spot in the biofilm (Figure 1E), which appears defined by the PI-labelling in the irradiation site. At the geometry and power used, most of the cells in the field of view are PI-labelled after blue-light irradiation (*Quantitative analysis to be published*). Neighboring non-irradiated biofilm is able to proliferate and our results show that upstream bacteria are also able to colonize the irradiated site, leading to the recovery/establishment of a biofilm structure post-irradiation (*data not shown*).

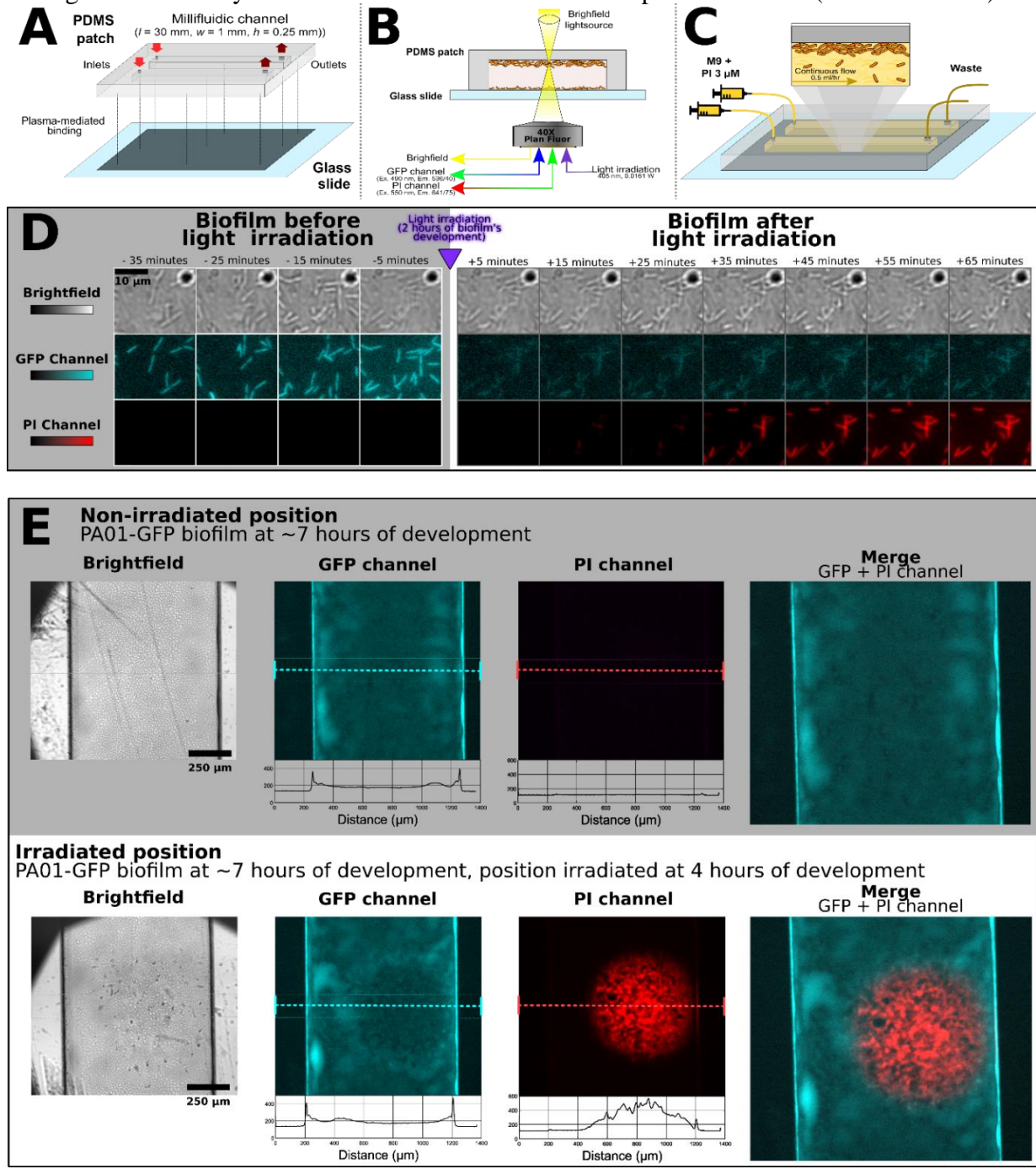

**Figure 1. A,** scheme of the plasma-mediated binding of the PDMS patch and the glass slide. **B**, scheme of the acquisition and irradiation conditions. **C,** scheme of PA01-GFP biofilm formation and constant washing of unbound cells. **D,** follow-up of bacteria behavior and development before (gray square) and after (white square) light irradiation at 2 hours of biofilm development (40X magnification: 40XSFluor (NA: 0.9, WD: 0.3 mm)). **E,** 4X magnification (4XPlan (NA: 0.1, WD: 30 mm)) of the millifluidic channel, in non-irradiated position (gray square) and an irradiated position (white square) at 4 hours of development; the fluorescence intensity was measured along a line (dotted line).

# 5 Conclusion

This work presents for the first time the combination of a millifluidic device and blue-light irradiation, aiming for the disinfection of pathogenic biofilm. The devised setup enables to deliver a spatially defined light dose, enabling users to deliver light at precise timepoints. The real-time microscope monitoring allows to follow up the biofilm development at cellular and community levels, enabling us to observe the dynamics that follow up after light irradiation, a phenomena yet to explore. Now, it is possible to tune and analyze the importance of different parameters (i.e., thickness of the biofilm, length of the irradiation, power used for the irradiation, geometry of the irradiation spot), parameters which most of the time are neglected and/or impossible to control in traditional approaches. The number of applications of this device are likely to grow with time, enabling to assess different light sources and exogenous photosensitizers, as an example.

# 6  *Disclosures:*

The authors declare no conflict of interest.

# 7  *Acknowledgments*

This research is part of the Light4Lungs project, which has received funding from the European Union's Horizon 2020 research and innovation programme under grant agreement nº 863102.

*References*

1. Verderosa, A. D., Totsika, M. & Fairfull-Smith, K. E. Bacterial Biofilm Eradication Agents: A Current Review. *Front. Chem.* **7**, (2019).
2. Martegani, E., Bolognese, F., Trivellin, N. & Orlandi, V. T. Effect of blue light at 410 and 455 nm on Pseudomonas aeruginosa biofilm. *J. Photochem. Photobiol. B* **204**, 111790 (2020).
3. Benyoussef, W., Deforet, M., Monmeyran, A. & Henry, N. Flagellar Motility During E. coli Biofilm Formation Provides a Competitive Disadvantage Which Recedes in the Presence of Co-Colonizers. *Front. Cell. Infect. Microbiol.* **12**, (2022).
4. Davies, E. V. *et al.* Temperate phages both mediate and drive adaptive evolution in pathogen biofilms. *Proc. Natl. Acad. Sci.* **113**, 8266–8271 (2016).
5. M9 minimal medium (standard). *Cold Spring Harb. Protoc.* **2010**, pdb.rec12295 (2010).
6. Treghini, C., Dell'Accio, A., Fusi, F. & Romano, G. Aerosol-based antimicrobial photoinactivation in the lungs: an action spectrum study. *Photochem. Photobiol. Sci.* **20**, 985–996 (2021).
7. Nikparvar, B. *et al.* A Diffusion Model to Quantify Membrane Repair Process in Listeria monocytogenes Exposed to High Pressure Processing Based on Fluorescence Microscopy Data. *Front. Microbiol.* **12**, (2021).